# A low-cost spectrum analyzer for trouble shooting noise sources in scanning probe microscopy

Nicholas M. McQuillan, Amanda M. Larson, and E. Charles H. Sykes
Department of Chemistry, Tufts University, Medford, MA 02155, USA
E-mail: Charles.Sykes@tufts.edu

**Abstract**
Scanning probe microscopes are notoriously sensitive to many types of external and internal interference including electrical, mechanical and acoustic noise. Sometimes noise can even be misinterpreted as real features in the images. Therefore, quantification of the frequency and magnitude of any noise is key to discovering the source and eliminating it from the system. While commercial spectrum analyzers are perfect for this task, they are rather expensive and not always available. We present a simple, cost effective solution in the form of an audio output from the instrument coupled to a smart phone spectrum analyzer application. Specifically, the scanning probe signal, e.g. the tunneling current of a scanning tunneling microscope is fed to the spectrum analyzer which Fourier transforms the time domain acoustic signal into the frequency domain. When the scanning probe is in contact with the sample, but not scanning, the output is a spectrum containing both the amplitude and frequency of any periodic noise affecting the microscope itself, enabling troubleshooting to begin.

**Introduction**

In recent decades, Scanning Probe Microscopes (SPMs) have experienced rapid development and employment after the invention of the Scanning Tunneling Microscope (STM)[1] and the Atomic Force Microscope (AFM).[2] The ability to measure physical and chemical properties of surfaces at the nanometer scale has been invaluable and has contributed to the widespread use of SPMs in fields such as chemistry, physics, materials science and even biology.[3–5] In general, these techniques rely on the use of a sharp probe tip, which in some cases is terminated by a single atom or molecule at the tip apex,[6] in close proximity to a sample surface. Piezoelectric actuators allow for the probe tip to be reproducibly moved very small distances in the x, y, and z-directions, thereby building a three-dimensional image of the sample surface.[3,6] Various feedback mechanisms are utilized in order to maintain the distance between the probe tip and sample (often referred to as the z-gap) constant in order to maintain for example, a constant tunneling current in STM, or force in AFM.[3,6,7] Given the scale at which these microscopy techniques work, the signals involved are extremely small. For instance, the STM measures tunneling current at the pA level,[8] while AFM relies force feedback as small as the pN scale.[9] Therefore, scanning probes often have problems with interference from external noise sources, which can skew results and produce surface "features" that can be misperceived as real.[10] Knowledge of the frequency of these external noise sources is crucial for finding their origin in order to eliminate noise.[11] Spectrum analyzers are very useful for the measurement of both the magnitude and frequency of noise sources in SPM images,[12,13] but in general they tend to be relatively expensive.

We describe an apparatus that just requires the use of a smart phone for determining the frequency of noise sources via a Fast Fourier Transform (FFT) application. Specifically, the SPM output signal, which in our case was the tunneling current from a STM, is connected to an amplified loudspeaker input. Several smart phone applications that can turn the acoustic signal from the speaker into frequency domain via an FFT plot are available. As many sources of noise for SPMs are low frequency, we tested the lower limit in terms of the frequencies our loudspeaker could produce. This involved connecting the speaker to another smart device with a noise generating application and testing to see whether low frequency signals could be transmitted via the speaker and detected in the spectra generated by the FFT application. We use several STM imaging experiments to demonstrate the capabilities of the setup in identifying the frequency of external noise sources.

**Results and Discussion**

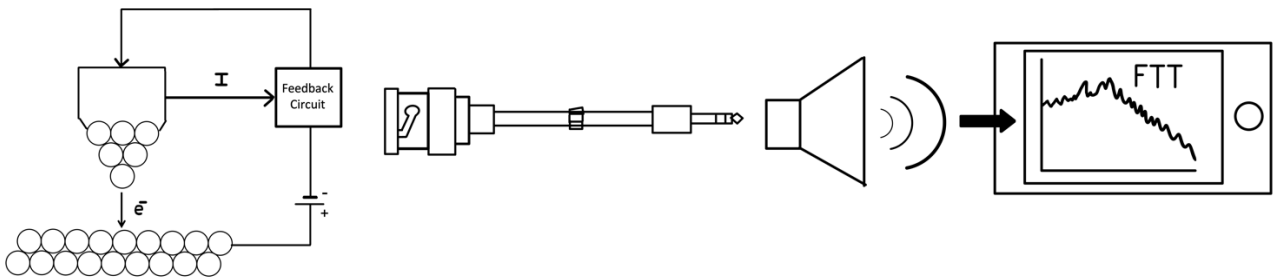

**Figure 1.** Schematic view of scanning probe microscope noise analysis apparatus. The STM tunneling current signal (or other microscope output) is converted into an audio signal with a loudspeaker coupled to a smartphone microphone, which via a FFT application converts to the frequency domain revealing the frequency of external noise sources.

The overall schematic of the noise analyzer setup for our STM experiments is outlined in Figure 1. As the scanning probe tip passes over a sample surface to which a positive voltage has been applied, electron tunneling takes place from the tip to the sample, producing an electron tunneling current that is measured. This tunneling current can be output to produce an image of the apparent topography of the surface, and in our case, to also produce the acoustic signal necessary for our smart phone spectrum analyzer input. This step was accomplished by fashioning a cable capable of converting the STM signal output to a loudspeaker input, which involved soldering a BNC connector cable and an audio jack cable together. The microscope output signal is converted acoustic noise and can be picked up by the built-in microphone of the smart phone and converted to a FFT plot in real time. We demonstrate this setup with the "FFT Plot" application by ONYX Apps, but there are several other options available that can perform the same function. The spectrum analyzer application measures the magnitude of the signal as sound pressure level (SPL) in decibels relative to full scale (dBFS), as well as the frequency in Hz. When analyzing a typical FFT spectrum of the STM tunneling current signal, there are typically several distinct peaks, the frequency of which enables the identification of sources of periodic noise.

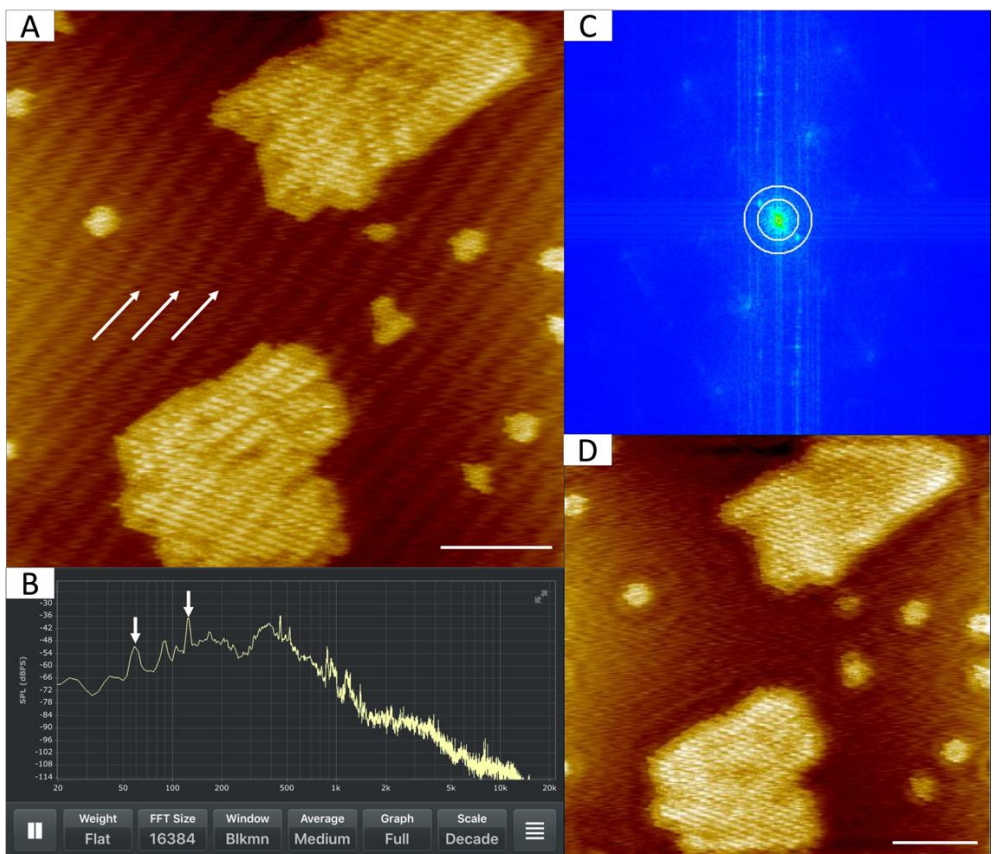

**Figure 2.** Demonstration of conversion of STM tunneling current signal into a FFT spectrum. (A) STM image of islands of 1,1,1-trifluoro-3-iodopropane molecules on a Cu(111) surface at 80 K. The brighter islands of molecules are real features, while the diagonal stripes highlighted by the white arrows arise from external noise. Scale bar: 10 nm. Imaging conditions: 80 mV, 220 pA. (B) An FFT plot of the STM tunneling current with highlighted peaks corresponding to the noise frequency (60 Hz, 120 Hz). (C) A 2D Fourier transform (FT) of the STM image in (A) with the 60 Hz noise features highlighted between the two white circles and this region was removed in order to produce the real space image in (D). (D) The same STM image as in (A), but with the 60 Hz noise removed via the 2D FT filter.

An example of an STM image with both real, and externally derived noise features is shown in Figure 2A. In this experiment, 1,1,1-trifluoroiodopropane molecules were deposited on a flat Cu(111) surface and imaged by STM. The large bright islands represent real features, i.e. semi-ordered, one molecule thick islands on the Cu(111) surface. The diagonal stripes (highlighted by white arrows in Figure 2A) imply the existence of alternating ridges and valleys on the Cu surface, but we know that this Cu surface is flat based on its (111) facetted surface and the fact that we have previously imaged the same system without these stripes present in a noise free environment. In order to characterize this particular noise feature and verify the corresponding features in the FFT spectrum, we first attempted to find the frequency of the noise from the image itself. This was done by using the known scan speed (in nm/s) and image size (in nm) and determining the period of the noise in a single linescan from the STM image. In the linescan each peak and trough are one waveform, and hence the number of waveforms per line can be calculated. This then allowed us to calculate the frequency of the noise, which in this experiment was 60 Hz. This frequency, as well as the first harmonic at 120 Hz, were clearly visible in the smart phone FFT spectrum (Fig 2B) verifying that the FT plot can detect noise sources in the image via the electron tunneling current signal.

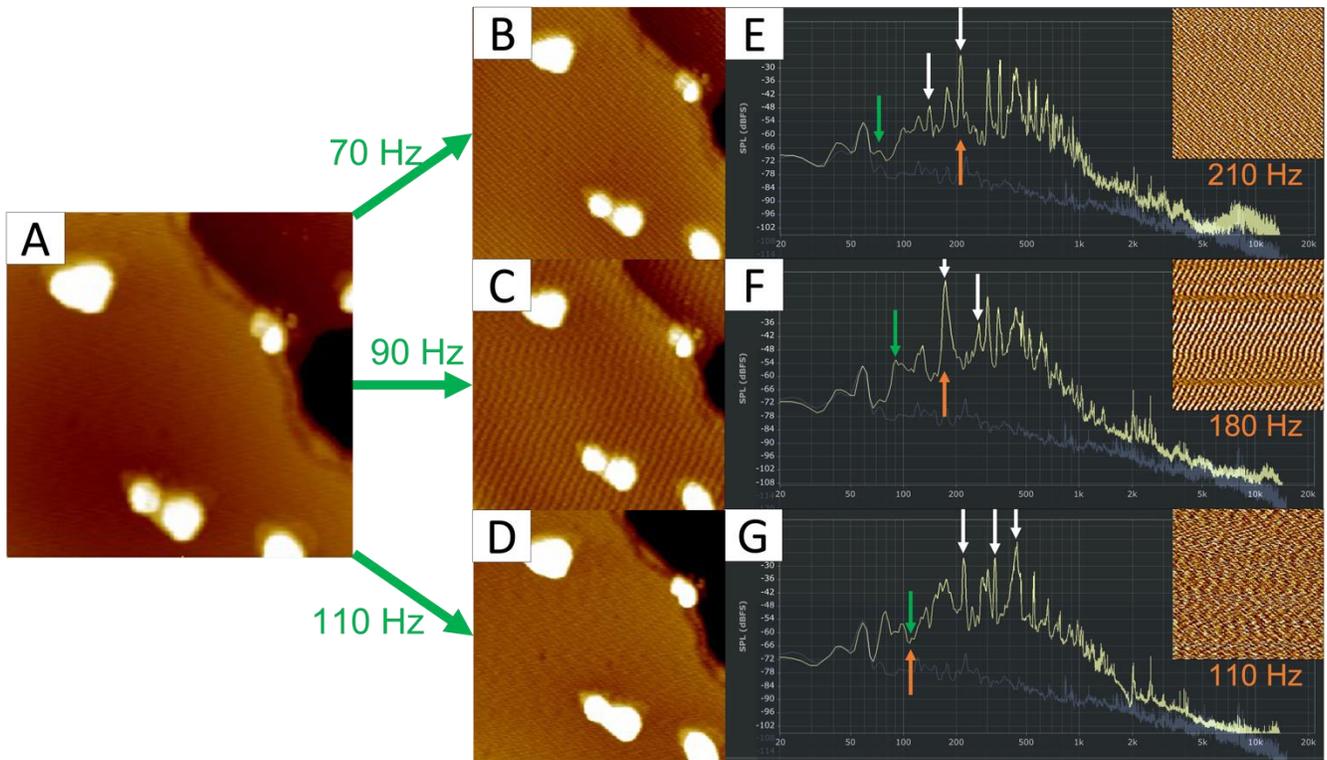

**Figure 3.** FT spectra of the STM tunneling current (yellow spectra) from several experiments in which noise signals with known frequencies were generated near the STM instrument, compared to background noise (blue spectra). The original STM image is seen without any noise (A) and STM images (B, C, and D) with added noise at 70 Hz, 90 Hz, and 110 Hz respectively. The corresponding FT spectra are shown in E, F, and G. STM images with the probe in point mode at constant current are shown in each respective inset. The orange arrows below the yellow FT spectra correspond to the noise frequency detected within the inset images using the method in the main text discussion of Figure 2A.

In order to test the ability of our setup to detect noise of unknown frequencies, we introduced acoustic noise of known frequencies beside the microscope via a signal generator. The STM itself was housed in a custom build "quiet room" with the control electronics outside the enclosure so that the only way these frequencies would be detected by our setup was via mechanical coupling of the generated noise to the ~1 nm STM tip-sample junction and subsequent coupling to the tunneling current signal. In Figure 3, the FT spectra of raw STM data measured via point mode in which the STM tip remains fixed are displayed. As can be seen in the figure, the intentional addition of different frequencies can be seen in the raw STM images (B, C, D), and in the FT spectra (E, F, G). Our apparatus could detect the fundamental driving frequency (green arrows) when the original noise signal was greater than 60 Hz, and in all cases could detect the first, second, and sometimes third harmonic frequencies (denoted by white arrows).

## Conclusions

We have demonstrated the capability of our simple homebuilt setup to identify the frequency of periodic noise present in our STM images as a result of external noise sources. This provides a fast and simple way to improve the quality of SPM images by identification of the frequency of the external interference which is crucial for identifying its origin. The primary limitation of this setup, as evidenced by our experiments adding noise with known frequency is that the apparatus can detect the fundamental frequency of the external source only if it is above ~60 Hz. This is not a major issue as below this limit, harmonic frequencies (for example 120, 180 Hz) can still be used to identify the external noise source. This apparatus can serve as a cost-effective alternative to using expensive commercially available spectrum analyzers in the trouble shooting of external sources of noise in many scanning probe techniques.

**Acknowledgements**
The project was supported by the US National Science Foundation under grant CHE-1764270.

**Data Availability Statement**
The data that support the findings of this study are available from the corresponding author upon reasonable request.